\documentclass[twocolumn,epsfig,prl,showpacs]{revtex4-1}

\usepackage{amsmath}
\usepackage{physics}
\usepackage{color,soul}
\usepackage{kotex}
\usepackage{ulem}

\usepackage{graphicx}
\usepackage{subfig} 
\usepackage{floatrow}
\usepackage[justification=RaggedRight]{caption}

\thisfloatsetup{%
objectset=raggedright,
}

\usepackage{multirow}   
\usepackage{tabu}
\usepackage{tabularx}
\usepackage{adjustbox}

\begin{document}
	\title{Quantified Degeneracy, Entropy and Metal-Insulator Transition in Complex Transition-Metal Oxides
	}
	
	\author{Jae-Hoon Sim$^1$}
	\author{Siheon Ryee$^1$} 
	\author{Hunpyo Lee$^2$}
		\author{Myung Joon Han$^{1,3}$} \email{mj.han@kaist.ac.kr} 
	
	\affiliation{$^1$Department of Physics, Korea Advanced Institute of Science and Technology (KAIST), Daejeon 34141, Korea}
	\affiliation{$^2$Department of General Studies, Kangwon National University, Samcheok-si, Kangwon-do 25913, Korea}
	\affiliation{$^3$KAIST Institute for the NanoCentury, Korea Advanced Institute of Science and Technology, Daejeon 34141, Korea}

\date{\today}






\begin{abstract}
Understanding complex correlated oxides and their phase transitions has long been a
challenge. The difficulty largely arises from the intriguing interplay
between multiple degrees of freedoms.  While degeneracy can play an
important role in determining material characteristics, there is
no well-defined way to quantify and to unveil its role in 
real materials having complicated
band structures. Here we suggest a way to quantify the `effective
degeneracy' relevant to metal-insulator transition
by introducing entropy-like terms. This new quantity well
describes the electronic behaviors of transition-metal oxides as a
function of external and internal parameters. 
With $3d$ titanates, $4d$ ruthenates,
and $5d$ iridates as our examples, we show that this new
effective quantity provides useful insights to understand these systems
and their phase transitions. For
LaTiO$_3$/LaAlO$_3$ superlattice, we suggest a novel `degeneracy
control' metal-insulator transition.
\end{abstract}

\maketitle

{\it Introduction}
Understanding transition-metal oxides (TMO) and their phase transitions
has been a central issue in condensed matter physics and material
science. Many of exotic quantum phases of matters
are the result of intriguing interplay and competition between
the multiple degrees of freedom active in TMO; namely, charge, spin,
orbital, and lattice \cite{Imada_RMP98}.
Estimating the key parameters which represent 
those degrees of freedom and their couplings is therefore
a crucial step. Quantifying
other physical parameters such as bandwidth, crystal field, 
and `interactions ({\it i.e.}, $U$, $U'$ and $J$)' is also
often posing a non-trivial task.
Depending on which parameter is crucial, the
metal-insulator transition (MIT) is
described and
classified into sub-category of `bandwidth control',
`filling control' and `dimensionality control' MIT
\cite{Imada_RMP98}.

Orbital degeneracy can certainly play an important role in
determining material characteristics of TMO \cite{Pavarini_PRL04,Zhong_Held_PRL15,Georges_AnnuRev13,Hossain_PRB_2012,Qi_PRL_2010,Brzezicki_PRX_2015}.
However, there is
no well-defined way to quantify and unveil its role in the coorperation with
other physical components of real materials
with complicated multi-band structures around Fermi energy ($E_F$).
In this study, we first try to quantify the `effective' orbital degeneracy
by introducing entropy-like terms.
Then we apply it to real material systems. Our
results of $4d$ ruthenates and $5d$ iridates show that this
newly-introduced quantity well describes the electronic behavior
and provides useful insight to understand MIT. Further, we
suggest a novel `degeneracy control' MIT in $3d$ titanate
superlattice. The strain-dependent phase diagram and the calculated
physical parameters clearly show that the transition is governed
mainly by `degeneracy' not by other factors such as bandwidth.

Defining an intuitive and computable physical quantity has been
playing central roles in quantum material research.
`Charge-transfer energy' for TMO \cite{Zaanen_Sawatzky_Allen_PRL85}
and 
`Chern number' (or TKNN number) for topological materials
 \cite{Thouless_TKNN_PRL82,Hasan_Kane_RMP10, SCZhang_RMP11}
can be recent examples. Even
though both are not directly measurable in experiments, they certainly
provide key information to classify and understand a given type of
materials. In this regard, the quantified `effective' degeneracy
we suggest here can also be a useful tool to study multi-orbital 
complex oxides and their phase transitions.

{\it Quantifying Effective Degeneracy}
In model-based studies, the degree of degeneracy is naturally defined
by the energy level difference. For real materials, on the other
hand, quantifying degeneracy is not always straightforward
due to the complicated band structures which is in the end
a result of combinations of many other `parameters' such as
crystal field levels and hybridizations. 
Furthermore we note that the information
relevant to MIT is hardly extracted from the `bare'
degeneracy represented simply by level differences.
With this motivation we define the following quantity:
\begin{equation}
D= \sum_\mu S(n_\mu).
\label{eq:D}
\end{equation}
Here the entropy-like term $S$ is given by
\begin{equation}
S(n_\mu)=-n_\mu\log_2 n_\mu -(1-n_\mu)\log_2 (1-n_\mu)
\label{eq:S}
\end{equation}
where $n_\mu$ is the eigenvalue of on-site number operator $\hat{N}$. The matrix elements of $\hat{N}$ can be written as $N_{\alpha\beta}=\frac{1}{N_{\bf k}} \sum_{\bf km} \braket{{\bf k} m}{\alpha}
\braket{\beta}{{\bf k} m}$ with orbital indices $\alpha$ and
$\beta$ ({\it i.e.,} three $t_{2g}$ states in our examples; well represented
by Wannier functions), momentum ${\bf k}$, and band index $m$. 
Note that $N_{\alpha \beta}$ is calculated from the `non-interacting' Hamiltonian, namely, $U=0$ paramagnetic band structutre,
which is the same case with other model parameters to be used to understand MIT such as bandwidth ($W$) and correlation strength ($U$) \cite{Aryasetiawan_PRB_2004, Jang2016}.
While the calculation of $D$ at finite $U$ is straightforward, the useful information
is mainly contained in $D$ at $U=0$.
{The eigenstate $\ket{\mu}$ of the on-site number operator} does not need to be any of conventional symmetry states ($\ket{\alpha}$), and can be expressed as the superposition of them \cite{Suppl}. 
$S$ is maximized ($S=1$) at half-filling ($n_{\mu}=0.5$) and minimized ($S=0$) when the orbital is fully
occupied ($n_{\mu}=1$) or unoccupied ($n_{\mu}=0$). 
$D$ reflects the number of orbital states at $E_F$ and thus carries similar information with degeneracy. 
At the same time, $D$ is clearly different from `mere' degeneracy.
The number of states is weighted by taking the band position into account with respect to $E_F$.
As a result, $D$ magnifies the information relevant to MIT.
In this sense, $D$ can be called as `effective degeneracy'. On the other hand, as obvious from Eq.~(2), $D$ measures a certain type of entropy being regarded as `effective entropy'.
For more discussion of its physical meaning, see Supplemental Material  \cite{Suppl}.

\begin{figure}
	\begin{center}
		\includegraphics[width=1.0\linewidth,]{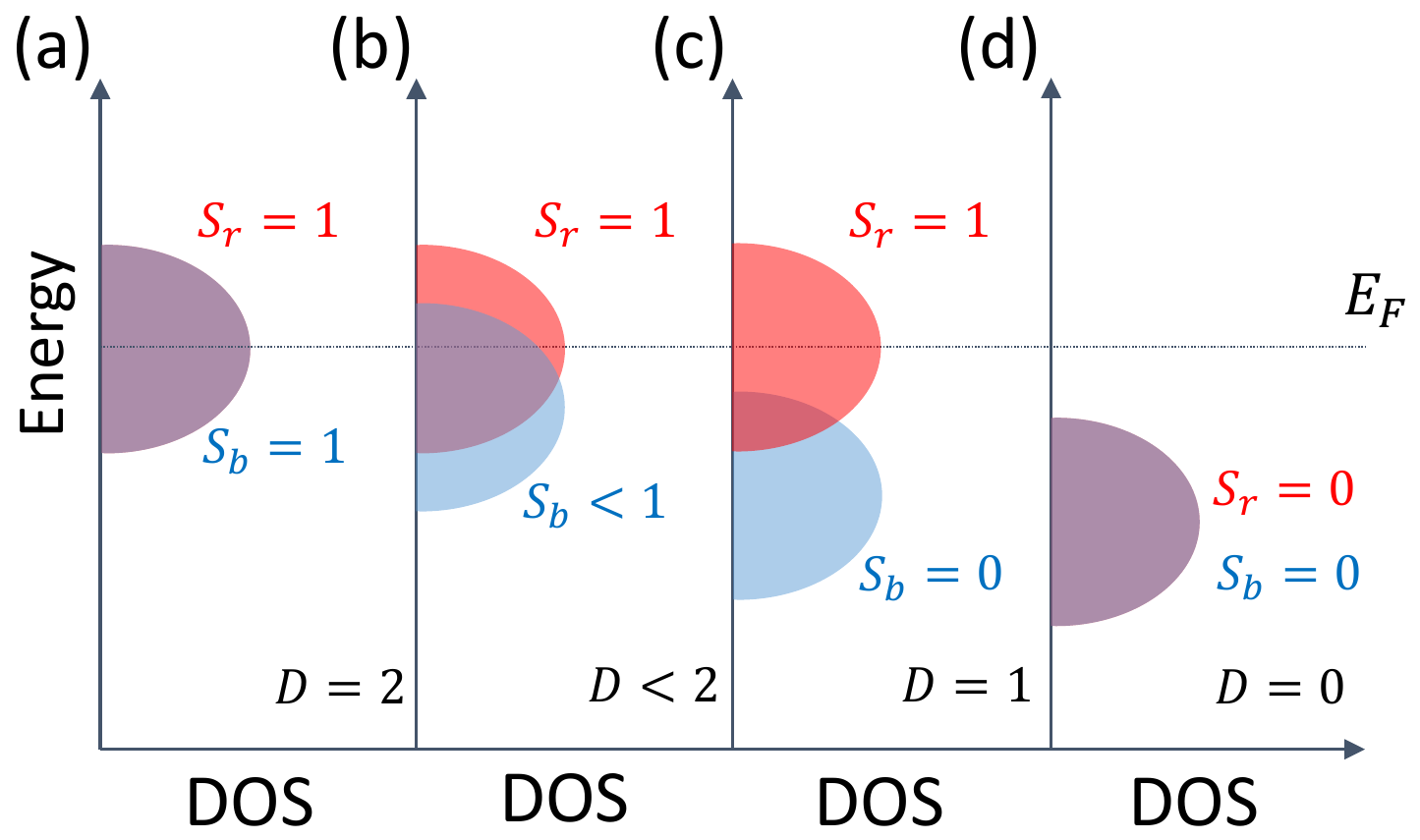}
	\end{center}
	\caption{
	    The behavior of  $D$ for some schematic
		two-orbital electronic structures.
        In the case of (a), two DOS are fully degenerated
        and $D$ is maximized, $D=2$.
    	As two DOS are separated from each other,
		$D$ is reduced as shown in (b) and (c).
		As an effective degeneracy (not `mere' degeneracy) 
		$D$ is being reduced when the on-site orbital energy 
		moves far away from $E_F$. As shown in (d), even if two DOS 
		are `fully degenerated' in the usual sense, 
		the calculated $D$=0 because both states are fully occupied
		\label{fig:schematic}
	}
\end{figure}

In order to see how this new quantity works, let us consider a
schematic situation presented in Fig.~\ref{fig:schematic}.
Physically,  degeneracy should be largest when the two bands
(assumed to have same bandwidths and shapes) are located at the same
energy; see Fig.~\ref{fig:schematic}(a). It is gradually lifted as
two levels become differentiated. Suppose that
the band structure evolves from Fig.1(a) to (b) and (c)
by any parameter change.
$S_{\rm b}$ (corresponding to one of the two bands; blue colored)
is gradually reduced from $S_{\rm b}$=1 (Fig.1(a)) to $S_{\rm b}$=0
(Fig.1(c))
while another band (red colored) does not move and 
$S_{\rm r}$ is unchanged.
When the two density of states (DOS) are identically overlapped with each
other (Fig.~\ref{fig:schematic}(a)), $D$ is maximized, $D=2$. When
the DOS overlap is minimized (Fig.~\ref{fig:schematic}(c)), $D$ is
minimized, $D=1$. Therefore $D$ carries the similar information with degeneracy
in the usual sense. At the same time, however, $D$
is different from `mere degeneracy' as clearly shown in
Fig.~\ref{fig:schematic}(d). 
The calculated $D$ of Fig.~\ref{fig:schematic}(d) is zero
even though two DOS are fully overlapped. This feature demonstrates
that $D$ is defined to represent the `effective' degeneracy or effective entropy
of the bands near $E_F$ relevant to MIT.
In fact, both DOS in Fig.1(d) are not relevant to MIT since they
are fully occupied and far away from $E_F$.

$D$ works for more general cases.
Consider $M$ orbitals whose on-site energies
are given by $\varepsilon$'s near $E_F$ ({\it i.e.},
$|\varepsilon-E_F| \ll W$ (bandwidth)).  
For two states per orbital (occupied and unoccupied), the number of configurations is given by $\Omega \sim 2^{M}$ and $M \sim \log_2 \Omega$.
In the case of single electron per site, 
$D(M)=M\log_2 M - (M-1)\log_2 (M-1)$ from $n_\alpha=1/M$. While $D(M)$ is not equal to but less than $M$ 
({\it e.g.}, $D \sim 3.90$ for $M=6$), it provides an acceptable measure of degeneracy or entropy for this given model in a
general sense.

{\it Applications to Real Materials}
The usefulness of this new quantity can be more clearly seen with 
real examples. In the below, we take three different systems of
$3d$, $4d$, and $5d$ TMO in which the degeneracy is changed
by the internal as well as external parameters. Also, our example
sets include both strongly (titanate and iridate; Mott insulators)
and moderately correlated (ruthenate in bulk; metals) electron systems.
In the below, while we define the standard deviation ($\sigma$) of Gaussian fitted DOS as the bandwidth, we confirm that the use of Wannier function or Lorentzian fitting does not change any of our conclusions.

\begin{figure*}
	\begin{center}
		\includegraphics[width=1.0\linewidth,]{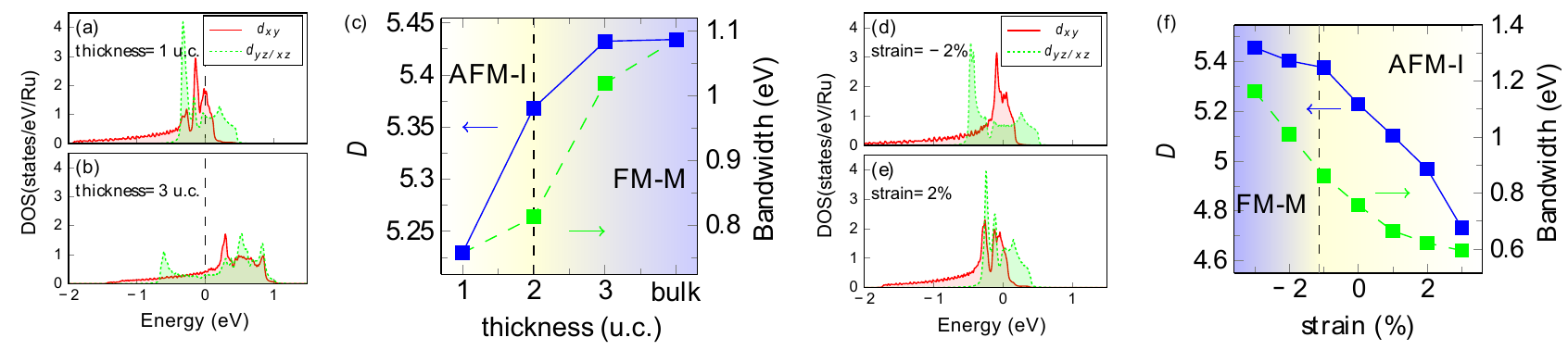}
	\end{center}
	\caption{ (a, b)
		The calculated DOS projected onto Ru-$t_{2g}$ states for 
		(a) $n$=1 and (b) $n$=3 with zero strain. The vertical dashed lines refer to $E_F$.
		(c) The calculated $D$ (blue; left side) and bandwidth (green dashed line; right side) as a function of film thickness, $n$. 
		A transition from FM metal (FM-M; blue-colored) to AFM insulator (AFM-I; yellow-colored)
		occurs at the thickness of $n$=2--4. (d, e)
		The calculated DOS projected onto Ru-$t_{2g}$ states for 1-unitcell-thick SRO-113 film with 
		(d) $-$2\% compressive and (e) +2\% tensile strain. 
		(f) The calculated $D$ and bandwidth as a function of strain. 
		At around $-1$\% compressive strain, there is a phase transition 
		from FM-M (blue-colored) to AFM-I (yellow-colored).
		The DOS, $D$, and bandwidth are calculated from non-spin-polarized ($U$=0) results while the phase diagrams are constructed from spin-polarized GGA calculations.
		\label{fig:ruthenates}
	}
\end{figure*}

{\it Ruthenates}
In this subsection we apply our `effective degeneracy' or `effective entropy'
to ruthenate thin films. Recently SrRuO$_3$ (SRO-113) has attracted
significant research attention due to its intriguing phase transitions
observed as the film thickness is reduced \cite{Toyota_APL05, Xia_PRB09,Koster_RMP12,Ishigami_PRB15,YJChang_PRL09}.
While bulk SRO-113 is
a ferromagnetic (FM) metal, its thin film phase is known to become 
insulating \cite{Toyota_APL05, Xia_PRB09,Koster_RMP12,Ishigami_PRB15} and presumably 
antiferromagnetic (AFM) \cite{Xia_PRB09,Si_PRB15,SRyee_SciRep17} . Critical thickness and the concomitance
of MIT and magnetic transition still remain unclear \cite{Toyota_APL05, Xia_PRB09,Koster_RMP12,Ishigami_PRB15,YJChang_PRL09}.

Fig.~\ref{fig:ruthenates}(a) and (b) shows 
the projected DOS for 1-layer ($n$=1) and 3-layer ($n$=3)
 film, respectively. As expected, the Ru-$t_{2g}$ states
are more degenerate in $n$=3 
being closer to 3-dimensional cubic bulk situation in which
the three are completely degenerate. On the other hand,
in $n$=1, $d_{xy}$ state  is noticeably different from
$d_{yz,zx}$.

Here we first note that this electronic structure change is 
 well described by $D$ in a quantitative manner.
As shown in Fig.~\ref{fig:ruthenates}(c), the calculated $D$ is 
gradually increased as $n$ increases being consistent with
the intuition and the band structure result.
It shows
that the newly-defined quantity, $D$, works reasonably well for
describing the realistic band structure change. Furthermore,
our result has a meaningful implication regarding the origin of 
observed MIT
as a function of thickness;
not only bandwidth (and/or dimensionality)\cite{YJChang_PRL09,SRyee_SciRep17}
but also degeneracy 
plays the role in this transition.

The usefulness of $D$ is further demonstrated in the
strain-dependent transition. Fig.~\ref{fig:ruthenates}(d) and (e)
presents projected DOS for 2\%
compressive and tensile strained mono-layer thin film,
respectively.
We note that in this case the degree of 
degeneracy is not clearly seen without calculating $D$.
The calculated $D$ as a function of strain is presented in Fig.~\ref{fig:ruthenates}(f) showing that $D$ is gradually
decreased as the more tensile strain is applied. 
In this sense the DOS of Fig.~\ref{fig:ruthenates}(d) is 
more `degenerate' than that of  Fig.~\ref{fig:ruthenates}(e).
This example
demonstrates that, even when the intuitive conclusion is not
likely reached, $D$ extracts the desired information.

Our result shows that $D$
is not `mere degeneracy' in the usual sense but reflects
the other factors relevant to MIT. According to
Eq.~(\ref{eq:D}) the states near $E_F$ 
contribute more to $D$ than the other states away from $E_F$.
Importantly, the decreasing trend of $D$
as a function of strain is consistent with the decreasing metallicity.
It is known that the system becomes more insulating in the tensile strain regime and more metallic in the compressive strain \cite{SRyee_SciRep17}.
In this regard, $D$ represents an `effective' degeneracy or entropy 
carrying the quantitative information physically relevant to
 MIT. In the case of DOS in Fig.~\ref{fig:ruthenates}(d),
the prominent $d_{xy}$ peak developed around $E_F$ is responsible
for larger $D$ (See Supplementary \cite{Suppl}).

{\it Iridates}
The second example is Sr$_2$IrO$_4$ (SIO-214) in which
the degeneracy is lifted not by external parameters
such as strain and thickness 
but by spin-orbit coupling (SOC). SIO-214 is
known as a `relativistic Mott' insulator in the sense that 
SOC plays the key role
to induce Mott gap \cite{BJKim_Science09,BJKim_PRL08}. Due to the large crystal field
and SOC, Ir-$t_{2g}$ states split into so-called
`$j_{\rm eff}$=3/2' quartet and `$j_{\rm eff}$=1/2' doublet.
Whereas different pictures have been discussed 
\cite{Arita_PRL12,Yamasaki_PRB14,Hongbin_PRL13,Hsieh_PRB12,Martins_PRL11,Sato_PRB15,Georges_JPIV04},
still quite prevailing is that the main role of SOC is to
reduce the bandwidth making relatively small $U$
be enough to open the gap \cite{BJKim_PRL08,Ishii_PRB11,Watanabe_PRL10}.

Here we show that the calculation of $D$ supplements
the understanding of Mott gap formation.
Let us first note that
SOC can not only reduce the bandwidth but also lift the degeneracy.
This possibility has been speculated in the context of 
reminiscing about multi-orbital
Hubbard model studies \cite{Martins_PRL11,Sato_PRB15,Georges_JPIV04}.
However, it could not be
discussed quantitatively and therefore not  
examined systematically 
in comparison to other possibilities 
\cite{Arita_PRL12,Yamasaki_PRB14,Hongbin_PRL13,Hsieh_PRB12}.
Further, a significant amount of hybridization between $j_{\rm eff}=1/2$ 
and $3/2$ has been noted in the literature \cite{Martins_PRL11, Watanabe_PRL10, Mohapatra_PRB_2017},
which render the simple model analysis more difficult.

\begin{figure}
	\begin{center}
		\includegraphics[width=1.0\linewidth,]{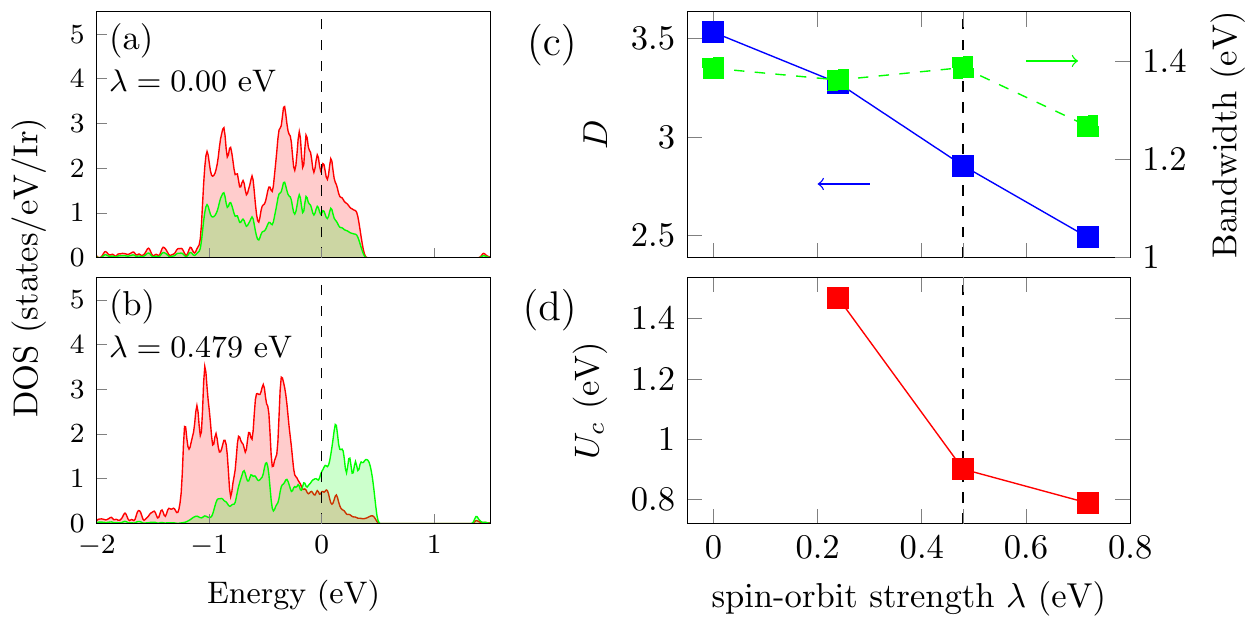}
	\end{center}
	\caption{(a, b) The calculated DOS projected onto so-called $j_{\rm eff}$ states
		(a) without and (b) with  SOC. The red and green colored states represent 
		$j_{\rm eff}=3/2$ and $1/2$, respectively. The vertical dashed lines correspond to $E_F$ 
		(c) The calculated bandwidth (green dashed line; right side) 
			of $j_{\rm eff}=1/2$ states
		and $D$ (blue; left side) as a function of $\lambda$. 
		Bandwidth are estimated from Gaussian fitting.
		The realistic value of $\lambda$=0.479 eV is denoted by the vertical dashed lines.
		(d) The calculated critical $U_c$ value as a function of $\lambda$.
		\label{fig:iridate}
	}
\end{figure}

Fig.~\ref{fig:iridate}(a) and (b) shows the projected DOS with
and without SOC, respectively, and the calculated bandwidth 
is presented in Fig.~\ref{fig:iridate}(c)  
as a function of SOC strength ($\lambda$).
In the range of $0\leq \lambda \leq 0.5$ eV,
the bandwidth is not significantly reduced. Rather,
our estimation shows a slight enhancement of bandwidth
at the realistic $\lambda$=0.479 eV.
In fact, the bandwidth reduction by SOC is not quite clearly
seen in the calculated DOS itself. 
Fig.~\ref{fig:iridate}(a) and (b) show that
the bandwidth of 
$j_{\rm eff}$=1/2 ($\lambda$=0.479 eV) can be regarded as being comparable 
with that of $t_{2g}$ ($\lambda$=0 eV).
Thus, it is difficult to conclude that the gap of SIO-214
is attributed to the bandwidth reduction by SOC.

Our new quantity provides useful insight on this issue.
The calculated $D$ is presented in Fig.~\ref{fig:iridate}(c)
whose decreasing trend clearly shows that the effective degeneracy is gradually lifted by SOC.
Together with the estimated critical $U_c$,  which follows the same decreasing trend
with $D$ (Fig.~\ref{fig:iridate}(d)), our results indicate that
lifting degeneracy is the main effect of SOC for the gap formation;
SIO-214 may still be regarded as a `relativistic Mott' insulator, but
the major role of SOC is not to reduce bandwidth but to lift 
degeneracy.

\begin{figure}
	\begin{center}
		\includegraphics[width=1.0\linewidth,]{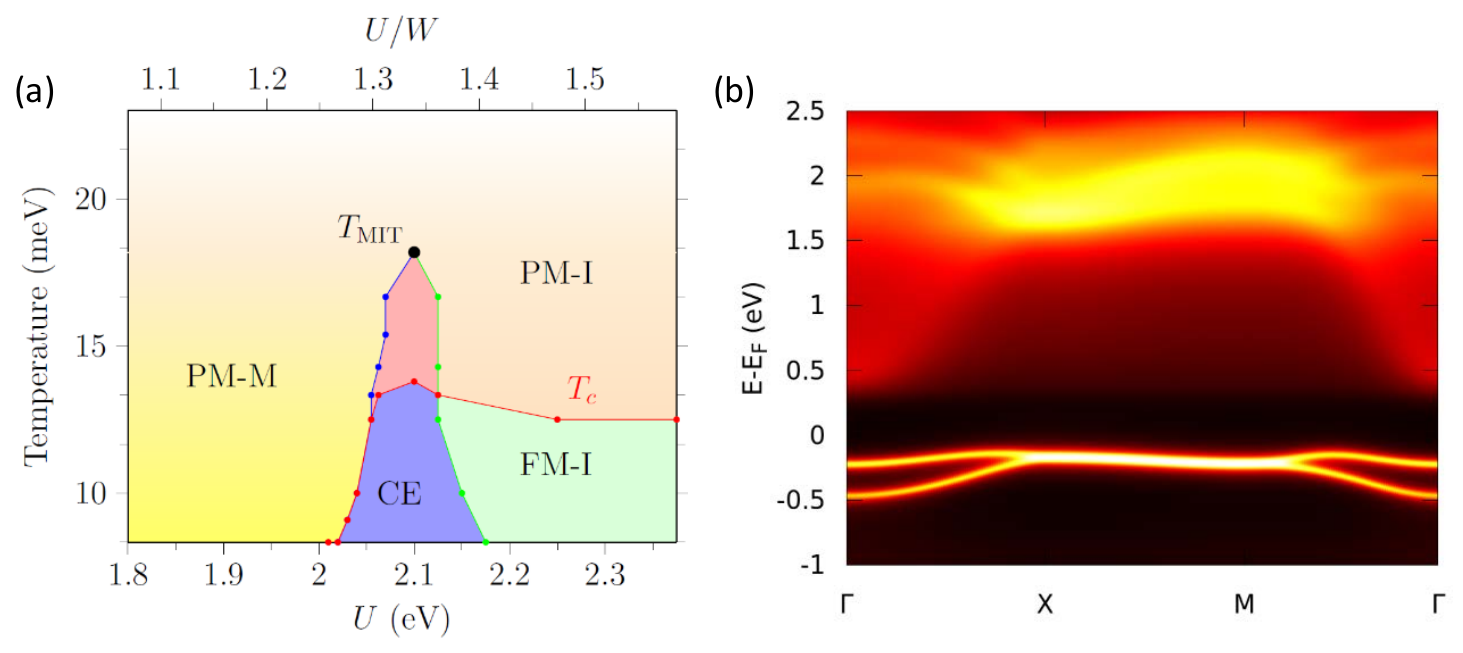}
	\end{center}
	\caption{ (a) The $U$-$T$ phase diagram.  The red line
          indicates the transition from the paramagnetic (PM) to the
          ferromagnetic (FM) phase.  The blue and green lines refer to
          the transition from metallic-to-insulating, and the
          insulating-to-metallic state, respectively. CE denotes the
          coexistence region indicating the first-order
          transition with an end point at $T_{\rm MIT}$. 
          (b) The calculated spectral function $A({\bf k},\omega)$
          projected onto Ti with $U=3 $ eV and at $T=8.3$
          meV.      
		\label{fig:PhaseDiagram}
	}
\end{figure}

{\it Titanates and degeneracy control MIT}
The final example is a superlattice made of a classical 
Mott insulator.  
The material property of LaTiO$_3$/LaAlO$_3$ (LTO/LAO) 
is an important issue by
itself since previous studies show that the electronic and
magnetic property are
notably different from bulk LTO due to confinement effect
\cite{SSASeoPRL10}.
Previous DFT$+U$ calculations
show that FM spin and antiferro orbital
order is stabilized \cite{ATLeePRB14}. However, no further
study has been reported especially using the more advanced techniques 
beyond static DFT$+U$, and a part of experimental observations
is still not clearly understood \cite{SSASeoPRL10}.
Our DMFT (dynamical mean-field theory) phase diagram is presented 
in Fig.~\ref{fig:PhaseDiagram}(a).
Paramagnetic insulating (PM-I) phase is clearly identified
at high temperature and large $U$ regime, which cannot be
addressed by static approximations. 
The calculated spectral function
$A({\bf k},\omega)$ is presented in
Fig.~\ref{fig:PhaseDiagram}(b). 
Coherent features is
noticed below $E_F$ \cite{Daghofer08PRL}
and the correct insulating nature is 
observed with 
$\Gamma$-point gap of 0.48 eV. This gap size is larger than
the bulk LTO optical gap ($\sim 0.2$ eV) being consistent with the
conductivity data on (LTO)$_{1,2,3}$/(LAO)$_5$
which reports that the lowest energy excitation is
gradually moving toward the higher frequency as the LTO layer thickness
is reduced \cite{SSASeoPRL10}.  
At a realistic value of $U=3$ eV for the superlattice \cite{ATLeePRB14,Weng_JAP15},
 the magnetic transition between ferromagnetic (FM-I) and 
PM-I occurs at $T_c \simeq 12.5$meV in good agreement with the total energy difference by GGA$+U$ \cite{ATLeePRB14}. Overall,
our DMFT results in large-$U$ and low-$T$ limit are consistent with static DFT$+U$ calculations. 

\begin{figure}
	\includegraphics[width=1.0\linewidth]{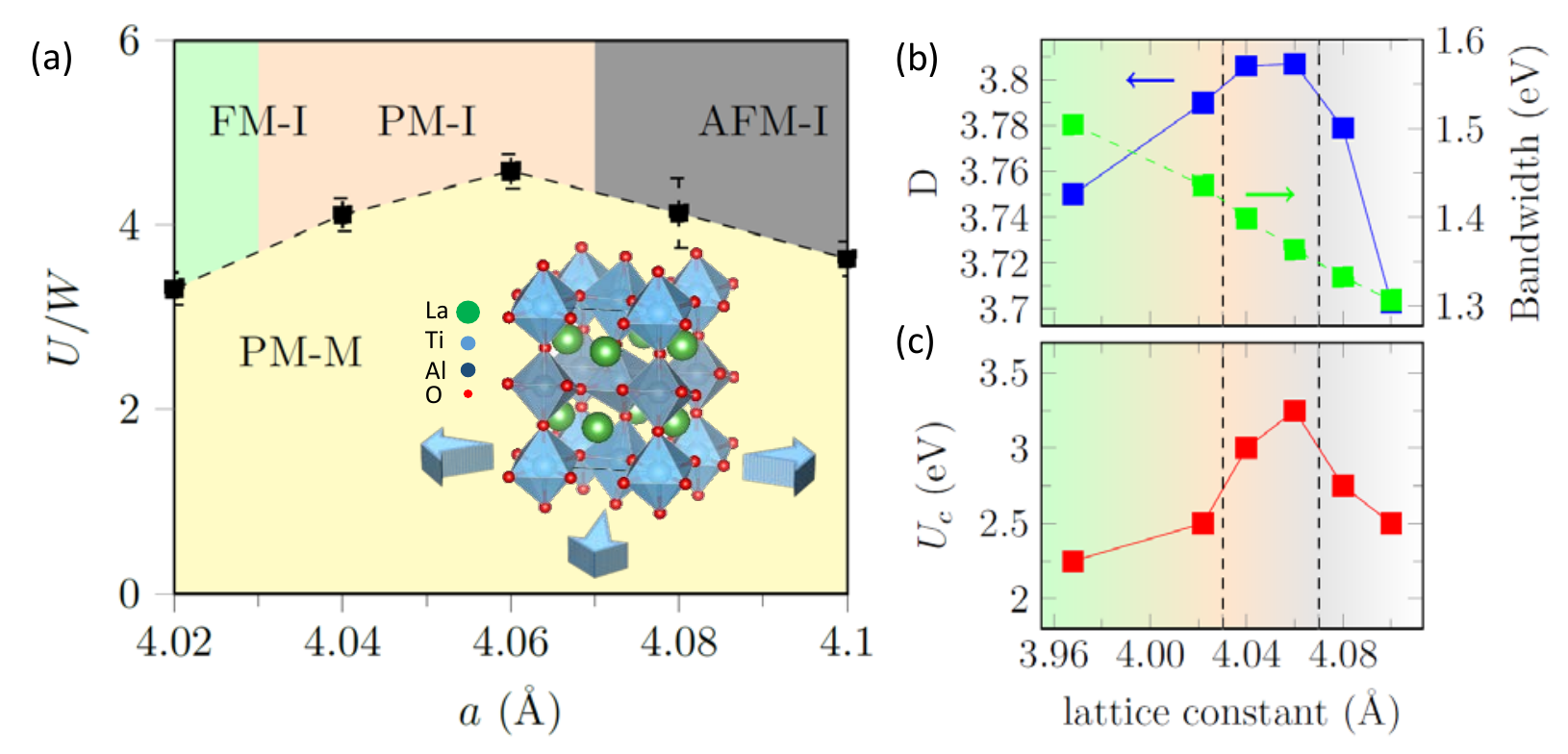}
	\caption{(a) Strain-dependent DMFT phase diagram in which
		the dashed line indicates the metal-insulator phase boundary.
          The metallic region (yellow colored) is enhanced at $a$=4.06\AA~ which
          corresponds to the degeneracy maximum point. 
          (Inset) Schematic figure for the `strain engineering' of
          LTO/LAO superlattice. (b) The calculated $D$ (blue; left side) and
          bandwidth (green dashed line; right side) as a function of in-plane lattice constant. As the
          more tensile strain applied the bandwidth is gradually
          decreased as expected. On the other hand, the $D$ is first
          increased and then decreased with a maximum value at
          $a$=4.06\AA. (c) The critical value of $U_c$ as a function of
          in-plane lattice parameter
          shows the same trend with $D$, suggesting
          that this phase transition is governed mainly by degeneracy.
		\label{fig:D_W_Uc}
	}
\end{figure}

Now we consider the strain dependence and `degeneracy control' MIT.
Fig.~\ref{fig:D_W_Uc}(a) presents the phase
diagram as a function of in-plane lattice parameter. Note
that metallic region (yellow) is enlarged at around
$a$=4.06{\AA}. 
The calculated $D$ is presented in Fig.~\ref{fig:D_W_Uc}(b)
showing that the effective degeneracy is also
maximized at this point.
The critical value of $U_c$
(Fig.~\ref{fig:D_W_Uc}(c)) exhibits the same trend.
Our results altogether indicate that the metallicity is
enhanced due to orbital fluctuations in the vicinity to
the degeneracy maximum point 
(For more details, see Supplemental Material \cite{Suppl}).

Importantly, this fluctuation overcomes
the effect of reduced bandwidth, thereby triggering the phase
transition.
The bandwidth is monotonically decreased; 
see Fig.~\ref{fig:D_W_Uc}(b), clearly indicating 
that the competition 
between metallic and
insulating phase is primarily governed by `degeneracy'.
In the current case, so-called `strain engineering' 
does not just control the bandwidth, but simultaneously change $D$,
and importantly, the governing parameter is (effective) 
degeneracy. Obviously, the other possibilities 
such as `dimensionality' and `filling' control MIT are not relevant
here.

As often being the case, this MIT is accompanied by
magnetic transition at low temperature. As shown in
Fig.~\ref{fig:D_W_Uc}(a), different magnetic phases are stabilized as a function
of in-plane lattice constant. While the
two end members of this phase diagram ({\it i.e.,} FM-I and AFM-I)
have been reported previously by GGA$+U$ calculation \cite{ATLeePRB14},
the PM-I phase at the degeneracy maximum point is first identified in 
the current DMFT study.

{\it Summary}
We introduced a new  measure of degeneracy, which estimates `effective' degeneracy
relevant to MIT. This quantity denoted by $D$ can be easily calculated
and is generally applicable to any
real multi-orbital systems for which quantifying
 the degree of degeneracy is often hampered by the 
 complicated band structures.
By applying $D$ to $3d$, $4d$, and $5d$ TMO,
we show that this newly-introduced quantity works well to describe the electronic behavior
as a function of external and internal parameters.
In particular, we show that the effective degeneracy
plays together with bandwidth change in the thickness-dependent
transition of SRO-113 and that 
the `relativistic' effect by SOC
in the SIO-214 gap formation is primarily to lift degeneracy rather than
to reduce bandwidth. From the strain dependent phase diagram of LTO/LAO superlattice, we suggest a novel `degeneracy control' MIT.
While `effective degeneracy' is not the only component to describe MIT, our examples clearly show that
this new quantity provides useful information which cannot be captured by other conventional quantities such as bandwidth and filling.

We thank A. T. Lee, H.-S. Kim, and S. W. Jang for 
 technical helps.  J.-H.S, S.R. and
M.J.H were  supported by
Basic Science Research Program through the National Research Foundation
of Korea (NRF) funded by the Ministry of Education 
(2018R1A2B2005204).
The computing resource is supported by
National Institute of Supercomputing and Networking /
Korea Institute of Science and Technology Information with supercomputing resources (KSC-2015-C3-042).




%

\end{document}


\title{Supplemental Material : Quantified Degeneracy and Metal-Insulator Transition in Complex Transition-metal Oxides}
	
	\author{Jae-Hoon Sim$^{1}$}
	\author{Siheon Ryee$^{1}$}
	\author{Hunpyo Lee$^2$}
	\author{Myung Joon Han$^{1,3*}$ \email{mj.han@kaist.ac.kr}}
	
	\affiliation{$^1$Department of Physics, Korea Advanced Institute of Science and Technology (KAIST), Daejeon 34141, Korea}
	\affiliation{$^2$Department of General Studies, Kangwon National University, Samcheok-si, Kangwon-do 25913, Korea}	
	\affiliation{$^3$KAIST Institute for the NanoCentury, Korea Advanced Institute of Science and Technology, Daejeon 34141, Korea}
	\date{\today}
\maketitle

\section{Computation method}
First-principles electronic structure calculations have been
carried out based on DFT (density functional theory) within 
GGA-PBE (generalized gradient approximation as parameterized by
Perdew-Burke-Ernzerhof) \cite{Perdew96PRL}. We used two
different software package, namely, 
`VASP' \cite{Kresse_PRB19_vasp} for ruthenates and titanates and
`OpenMX' \cite{MJHanPRB06,openmx,ozaki04PRB,ozaki03PRB} for
iridates and titanates. 
In SRO-113 calculation, 8$\times$8$\times$2 $\vb{k}$-points
have been taken and the slab geometries with a vacuum thickness of $\geq 15$ \AA~
considered. The force criterion for structural optimizations was 1 meV/\AA~.
Other details can be found in our previous publication of SRO-113 thin film \cite{SRyee_SciRep17}.
For SIO-214, we used DFT$+U$ method \cite{Dudarev98PRB, MJHanPRB06}
to take on-site correlation into account with 8$\times$8$\times$4 $\vb{k}$-points.
Critical value of $U_c$ was defined as the
smallest $U$ at which the gap is opened. SOC is treated within a fully relativistic $j$-dependent potential.
For LTO/LAO superlattice, we used 7$\times$7$\times$3 $\vb{k}$-points
and the optimized lattice structure obtained in our previous study \cite{ATLeePRB14}.
To describe the electronic correlation, single-site DMFT has been adopted.
Three maximally localized Wannier functions (WFs) are constructed
for each Ti site starting from the initial projections onto atomic Ti-$t_{2g}$
orbitals \cite{Mazari97PRB,Souza01PRB}. This Hamiltonian serves
as the non-interacting $H_0$ for the multiband Hubbard
Hamiltonian $H=H_0 + H_{\rm int}$. The interaction part is expressed
in the Slater-Kanamori form $H_{\rm int} =\sum_i h_{i,\rm int}$;
\begin{align}
h_{i,\rm int} &=  \sum_{\alpha} U n_{i\alpha\uparrow} n_{i\alpha\downarrow}  
+  \sum_{\alpha\neq \beta} U^\prime n_{i\alpha \uparrow }n_{i\beta \downarrow } \\
&+ \sum_{\alpha\neq \beta, \sigma} (U^\prime-J_H) n_{i\alpha \sigma }n_{i\beta \sigma },
\end{align}
where $U$, $U^\prime$, and $J_H$ refers to the intra-orbital, inter-orbital, 
and Hund interaction, respectively, with $U^\prime = U-2J$ and 
$J_H/U=0.14$ \cite{Pavarini_PRL04}. The Hamiltonian is
solved within single-site DMFT (dynamical mean-field theory) 
by employing a hybridization expansion
continuous-time quantum Monte Carlo (CT-QMC) algorithm implemented in
ALPS library \cite{ Alet_JPSJS05_ALPS2.0, Gull_RMP11,Gull_CPC11_CTQMC}.  In this procedure, the local
Green's functions is calculated
using momentum-independent self-energy:
\begin{equation}
G_{\rm loc}(i\omega_n) =  \frac{1}{N_k} \sum_{\bf k}  \frac{1}{i\omega_n +\mu - H_0({\bf k}) - \Sigma(i\omega_n)},
\end{equation}
where $H_0({\bf k})$ and $\Sigma(i\omega_n)$ are given by $12 \times 12 $
matrices. Self-energy is decomposed into $6\times6$
matrices corresponding to two Ti sites,
$\Sigma(i\omega_n)=\Sigma_{\rm Ti(1)}(i\omega_n) \oplus \Sigma_{\rm
	Ti(2)}(i\omega_n)$.
The self-energy in the real frequency domain is obtained from the
matsubara Green's functions and self-energy by analytic continuation
using the maximum entropy method (MEM) \cite{Jarrell96PhysRep,Gunnarsson_PRB10_MaxEnt}.
The stability is also checked by using stochastic
method.


\section{Supplementary Table: The calculated values of $D$ and $S_{n_\alpha}$}

\begin{table}[h]
		\caption{
		The calculated $D$ and $S_{\mu}$ for three different systems considered in this study. 
		Note that the states $\ket{S_{1,2,3}}$ are obtained from the diagonalization 
			of the number operators, and therefore do not exactly correspond to
			any of conventional symmetry states.
		The `control parameters' refers to the value of strain, the SOC strength ($\lambda$), 
		and the in-plane lattice constant for SRO-113, SIO-214, and LTO/LAO, respectively.
	}
\begin{adjustbox}{width=0.8\textwidth,center}
		\begin{tabular}{|c | c | c | c | c |   c |}
			\hline
        &control parameter  & $D$   & $S_1$ & $S_2$ & $S_3$    \\  \hline
\multirow{3}{*}{SRO-113}  & +2\%      & 4.949 & 0.941 & 0.987 & 0.494 \\   
                     &  0\%      & 5.255 & 0.977 & 0.974 & 0.676  \\ 
                     & -2\%     & 5.416 & 0.959 & 0.956 & 0.793  \\ \hline \hline
\multirow{2}{*}{SIO-214}  & 0.000 eV  & 3.490 & 0.784 & 0.782 & 0.179  \\ 
                     &0.479 eV  & 2.855 & 0.980 & 0.394 & 0.053  \\ \hline \hline
\multirow{4}{*}{LTO/LAO}  &3.905 {\AA}    & 3.751 & 0.806 & 0.663 & 0.405  \\ 
                     &4.040 {\AA}    & 3.781 & 0.780 & 0.650 & 0.441  \\ 
                     &4.060 {\AA}    & 3.805 & 0.787 & 0.636 & 0.478  \\ 
                     &4.100 {\AA}    & 3.697 & 0.857 & 0.547 & 0.445   \\ \hline
		\end{tabular}
	\end{adjustbox}
        \label{Table:Calculated_D}
\end{table}


\clearpage

\section{Supplementary Figure: Strain dependent DOS for LTO/LAO}

\begin{figure} [h]
	\includegraphics[width=0.8\linewidth]{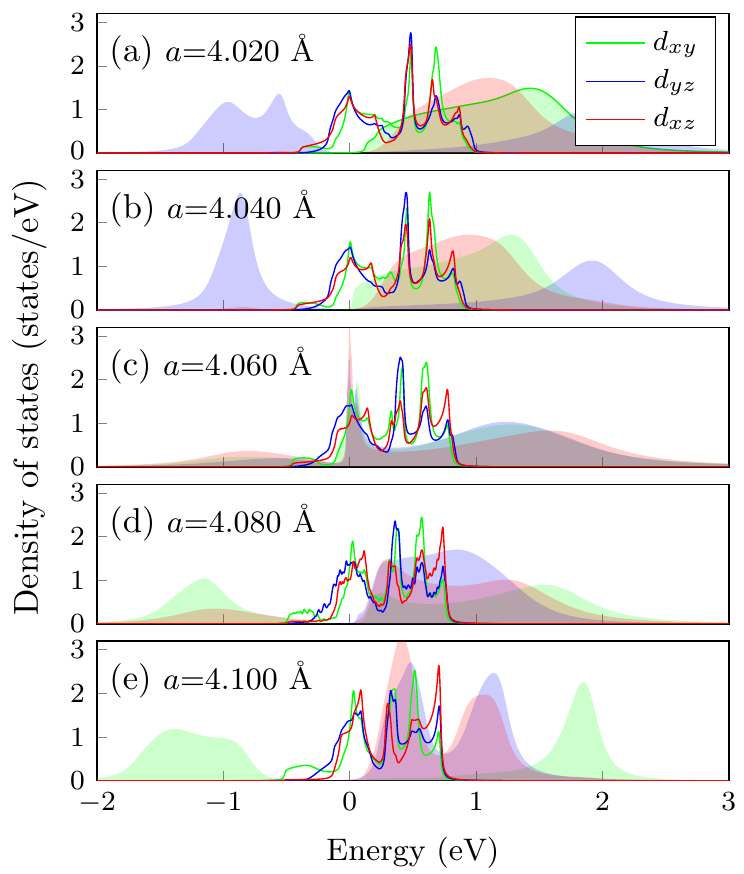}
	\caption{
		The orbital-resolved spectral functions of one of Ti sites 
		as a function of in-plane lattice constant. $U$=0 and $U$=3eV results are presented with
		the lines and the shaded area, respectively. The $d_{xy}$, $d_{yz}$ and $d_{zx}$ is represented by green, blue, and red colors, respectively. The Fermi level is set to 0.}
	\label{fig:dos_strain}
\end{figure}

\clearpage


\section{Physical Meaning of $D$}

$D$ is expressed with binary entropy of the eigenvalues of on-site number matrix,
$S(n_\mu) =-n_\mu \log _2 n_\mu -(1-n_\mu) \log _2 (1-n_\mu)$.
Since $n_\mu$ can be interpreted as a probability that the orbital $\mu$ is occupied, 
the probability $P_{\rm loc} (\{n_{\mu}\}) $ of the local micro-states $n_{\mu} \in \{ {\rm occupied} , {\rm unoccupied} \}$ can be written as 
$\Pi_{\mu}^{\rm occ} n_{\mu} \Pi_{\nu}^{\rm unocc} (1-n_{\nu})$. 
Here the `correlations' between electrons are not taken into accounted which is consistent with that
$D$ is calculated within the `non-interacting' Hamiltonian.
From the additivity of logarithmic function,
\begin{equation}
D= S(\Pi_\mu  n_\mu)= S(P_{\rm loc} (\{n_{\mu}\})) = S_{\rm vN}(\hat \rho_0),
\label{eq:qentropy}
\end{equation}
where 
$S_{\rm vN}(\hat \rho_0)=-\Tr (\hat \rho_0 \log \hat \rho_0 )$ is the von Neumann entropy for the quantum mechanical density matrix $\hat \rho_0 = \Tr_{i\neq0}  \ketbra{\Psi}{\Psi} $ at the local site $i=0$ and ground state $\ket{\Psi}$.
The relation between $D$ and the entropy (Eq.~(\ref{eq:qentropy})) is thus quite natural.

Note that the larger $D$ implies the larger fluctuation or entropy of a system.
As obvious from its entropy-like definition, $D$ measure the on-site electron fluctuations between different orbitals which suppresses the tendency \cite{Florens2002, Gunnarsson1996}.

Not surprisingly, the physical meaning of $D$ is multi-fold being related to
other more `conventional' quantities. As discussed in the main manuscript, $D$ carries
the information of band fillings as well as the 
(mere) degeneracy and entropy
relevant to metal-insulator transition \cite{Florens2002, Gunnarsson1996}.
 Further, $D$ can conceptually
be extended and closely related to `entanglement entropy' in a certain parameter regime \cite{Nielsen_Book2010}. It seems a promising future direction to explore the physical implication of $D$.


%